# Routing in Networks on Chip with Multiplicative Circulant Topology

**M A Shchegoleva**[1], **A Yu Romanov**[1,2,3], **E V Lezhnev**[1], **A A Amerikanov**[1]

[1] National Research University Higher School of Economics, 20 Myasnitskaya Ulitsa, 101000, Moscow, Russia
[2] e-mail: a.romanov@hse.ru
[3] Author to whom any correspondence should be addressed

**Abstract**. The development of multi-core processor systems is a demanded branch of science and technology. The appearance of processors with dozens and hundreds of cores poses to the developers the question of choosing the optimal topology capable to provide efficient routing in a network with a large number of nodes. In this paper, we consider the possibility of using multiplicative circulants as a topology for networks-on-chip. A specialized routing algorithm for networks with multiplicative circulant topology, taking into account topology features and having a high scalability, has been developed.

## 1. Introduction

Currently, one of the most important areas of research in the field of computer science and computing systems is the construction of multi-core processors. The transition to multi-core processors allows overcoming the performance degradation in the design of increasingly complex single-core systems [1]. When growing interest in the Systems on Chip (SoCs) and Multi-Processor Systems on Chip (MPSoCs) [2], Networks-on-Chip (NoCs) [3] are becoming widespread. In a multi-core processor with a small number of cores (2–8 cores), communication between IP-cores occurs via a common bus not capable to ensure communication between a large number of cores [1]. The problem of scalability can be solved by technology for NoC construction used to replace bus architectures.

One of the most urgent problems in the study of NoCs is the search for optimal topologies, since standard regular topologies (mesh [4], torus [5], hypercube [6], spidergon [7]) do not meet modern requirements for NoCs, especially with an increase in the number of nodes [8].

Circulant topologies have better characteristics than standard ones: they have better indicators of structural survivability, reliability, and connectivity, and also require fewer inter-processor exchanges in solving computing tasks, and system management tasks [9]. This allows them to be used in networks with a large number of nodes [10, 11]. To use circulant topologies as a topology for NoCs, it is necessary to develop routing algorithms in them taking into account the features of these families of graphs and the organization of NoCs, and all these determine the relevance of this work.

## 2. Multiplicative circulants

Let us give the definition of a circulant graph. Let $s_1, s_2, \ldots, s_k, n$ – whole numbers, such as $1 \leq s_1 < s_2 < \ldots < s_k < n/2$. Graph $C$ with set of vertices $V = \{0, 1, \ldots, n-1\}$ and set of edges $E = \{(i,j): |i-j| \equiv s_l \bmod n, \ l = \overline{1,k}\}$ is called circulant [12, 13]; numbers $S = (s_1, s_2, \ldots, s_k)$ – generatrices; numbers $k$ and $n$ – dimension and order of the graph [14].

Circulant graphs of $C(s^k; 1, s, s^2, \dots, s^{k-1}) = MC(s, k)$ with $c = 1$ and $n = s^k$ (Figure 1) are considered to be a separate class of multiplicative circulants [15].

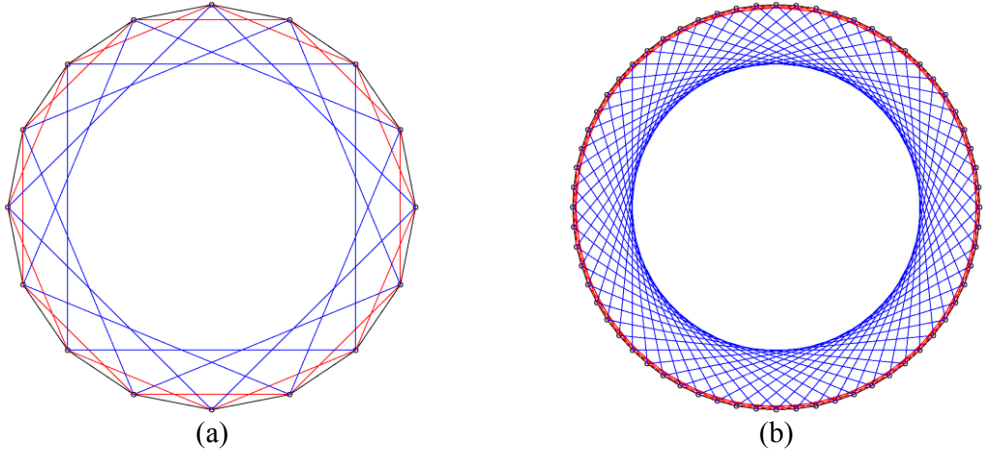

(a) (b)

**Figure 1.** Multiplicative circulant $MC(2, 4)$ (a) and $MC(4, 3)$ (b).

The most important characteristic of a graph is its diameter and average distance. The diameter of a graph $G$ is the maximum possible distance (the length of the shortest path) between two vertices.

$$d(G) = \max_{i,j \in V} d(i, j),$$

where $d(i, j)$ – the length of the shortest path from vertex $i$ to $j$.

In a number of works [15, 16], attempts have been made to find the most accurate estimates of the diameter and the average distance for multiplicative circulants.

More accurate estimates of the diameter (1) and mean distance (2) for the class of multiplicative circulants $MC(2, k)$ are given in [17]:

$$d(MC(2, k)) = \left\lceil \frac{k}{2} \right\rceil, \tag{1}$$

$$L_{av}(MC(2, k)) \approx \frac{k}{3}. \tag{2}$$

A comparison of the main characteristics of topologies on the basis of multiplicative circulants (1-2) and mesh-type topology with the same $n = s^k$ number of nodes is presented in Table 1.

**Table 1.** Comparison of characteristics of the circulant and mesh topologies.

| Circulant $MC(s,k)$ | Number of nodes $n$ | Diameter $d(MC)$ | Average distance $L_{av}(MC)$ | Diameter $d(mesh)$ | Average distance $L_{av}(mesh)$ |
|---|---|---|---|---|---|
| MC(2,4) | 16 | 2 | 1,33 | 6 | 2,50 |
| MC(2,6) | 64 | 3 | 2,00 | 14 | 5,25 |
| MC(3,4) | 81 | 4 | 2,67 | 16 | 5,93 |
| MC(5,4) | 625 | 8 | 4,80 | 48 | 16,64 |
| MC(3,6) | 729 | 6 | 4,00 | 52 | 17,98 |
| MC(6,4) | 1296 | 10 | (5,00, 6,00) | 70 | 23,98 |
| MC(7,4) | 2401 | 12 | 6,86 | 96 | 32,65 |

To calculate diameter and average distance of mesh topology, the formulas (3-4) [8, 18] are used.

$$d(mesh_n) = 2(\sqrt{n} - 1), \tag{3}$$

$$L_{av}(mesh_n) = \frac{2(n-1)}{3\sqrt{n}}, \tag{4}$$

From Table 1 it follows that even with a small network size, the circulant topology has better performance in all characteristics compared to mesh topology. Thus, multiplicative circulants are good enough for NoC designing with a large number of cores. However, it should be noted that the this topology, have a limited number of options – the number of nodes must be strictly an integer degree. Otherwise, the circulant becomes a recursive one with other properties and characteristics [19].

## 3. Development of the packet structure for static routing in networks with topology based on multiplicative circulants

Circulant networks use pairwise routing [14], when a packet is sent from the source node to the receiver node. For pair routing in NoCs with a multiplicative circulant topology, it is possible to use standard shortest path search algorithms, for example, breadth-first search algorithm (BFS) [20, 21]. It is suggested to choose a static type of routing in which each router has adjacency list for each node. Each router knows its own serial number, i.e. the node number whose incoming packets it distributes.

The destination node number (receiver node) is input to the router. The router calculates shortest path by the search algorithm in width. Then the router overwrites the path in the reverse and replaces the node numbers with the port numbers for which the packet is to be sent. A port is a connection between the current node and others. Each node has $2k$ ports: 2 ports of each length from S (except for circulants $MC(2, k)$, where one port is less). Thus, for circulant $MC(4, 3)$ (Figure 1b) the path $5 \rightarrow 21 \rightarrow 17$ is converted to $17 \leftarrow 21 \leftarrow 5$, then to the sequence of actions –(4), +(16), and then to 2|1 or 010|001 in binary code, if the rule for determining the port numbers is set as following: –(16) «left at 16»; –( 4) «left at 4»; –( 1) «left at 1»; +( 1) «right at 1»; +(4) «right at 4»; +(16) « right at 16».

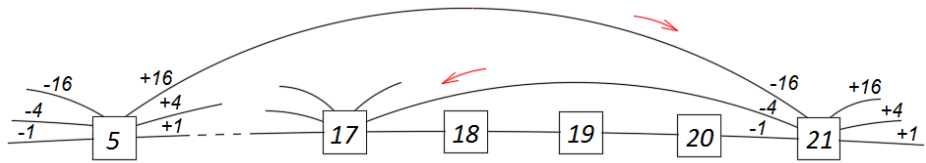

**Figure 2.** Path from node 5 to node 17 for circulant $MC(4,3)$.

Replacing node numbers to port numbers allows reducing size of part of the packet that is allocated for storing the path. Each node reads last k bits, shifts the path to same number of bits to the right, and passes the packet further. The package reaches the destination when the entire path is filled with zeros.

This algorithm is universal and it is suitable for networks built on the basis of any multiplicative circulants with difference $s$ and $k$. The problem with this algorithm is that with the increase in the number of nodes and connections, the running time of the algorithm is significantly increased too. Also, the size of the address part of the packet increases, so for large NoCs, it is required to develop a specialized algorithm optimized for this type of topology.

## 4. Development of a specialized routing algorithm for NoCs based on the topology multiplicative circulant

The review of algorithms of routing used in NoCs [22], shows that the most common in networks with a mesh-like topology is the XY algorithm [23]. Multiplicative circulants also have a strict, previously known geometric form. Therefore, if we take into account the peculiarity of this topology, consisting in the fact that the lengths of generatrices are powers of one base, it is possible to propose a specialized routing algorithm that simplifies the structure of the address part of the packet and reduces its size.

The address field stores the number of the destination node; the field size can be calculated by the formula $P = \lceil \log_2 N \rceil$, where $N$ – number of nodes in the network:

Receiving the node number where the packet is to be delivered, the router of the current node does not calculate all the way, but only the next step. In order to calculate the next step, it is enough for the router to know its own number, destination node number and circulant characteristics – $s$ and $k$.

Thus, the total size of the data, stored by the network routers in bits, can be calculated by the following formula:

$$M = N * (\lfloor \log_2 N \rfloor + \lceil \log_2 N \rceil + k * (\lfloor \log_2 S^{k-1} \rfloor + 1) + 3 * \lceil \log_2 k \rceil + 2), \qquad (6)$$

where $N$ – number of nodes in the network;
$\lfloor \log_2 N \rfloor + 1$ – required amount of memory (bit) to store the number of routers in the network;
$\lceil \log_2 N \rceil$ – required amount of memory (bit) to store the network router number;
$k * (\lfloor \log_2 S^{k-1} \rfloor + 1)$ – required amount of memory (bit) to store array of the generatrices;
$\lceil \log_2 k \rceil + 1$ – required amount of memory (bit) to store the indices of the generatrices and the primary port of the router and flag of primary port;

Taking into account the fact that the circulant topology is cyclic, the algorithm is developed from the position of the null node. To do this, before starting work, the destination node number is recalculated, based on the current node number: if the number of the current node is greater than the destination node number, the recalculated destination node number is equal to the difference of the two numbers. If the destination node number is less than the current node number, the resulting difference is subtracted from the number of nodes in the network to allow for the transition through the null node. Then the algorithm determines in which direction it is better to start moving – to the left, or to the right. Since the circulant is symmetrical, it is sufficient to consider half of the circulant and by fixing the chosen direction of motion. The step is determined by the length of the generatrix closest to the destination node number. This makes it possible to take into account the cases when it is more advantageous to overstep the destination node, and then go back. To do this, a generatrix with a maximum length, not exceeding the destination node number, and next longer generatrix are chosen. Of the two generators, it is selected the one whose absolute value of difference between its length and destination node number is the minimum. The packet reaches the destination node when the current node number is equal to the node number specified in the address part of the packet. Thus, for routing in the address part of the packet, it is necessary to provide $\lceil k \cdot \log_2 s \rceil$ bits for storing the destination node number.

The algorithm, specially developed for multiplicative circulants, requires significantly less time for calculations (Table 2) and is easily scaled to larger networks. This algorithm has a linear complexity in contrast to the width-search algorithm with an exponential dependence of the search path time on the graph dimensions.

**Table 2.** The time spent on searching for the shortest path in multiplicative circulants.

| Circulant | Number of nodes | BFS | Specially developed algorithm |
| --- | --- | --- | --- |
| $MC(2,4)$ | 16 | 0,000 | 0,000 |
| $MC(2,5)$ | 32 | 0,002 | 0,000 |
| $MC(2,6)$ | 64 | 0,013 | 0,001 |
| $MC(3,4)$ | 81 | 0,029 | 0,000 |
| $MC(5,3)$ | 125 | 0,340 | 0,001 |
| $MC(3,5)$ | 243 | 2,096 | 0,004 |
| $MC(6,3)$ | 216 | 10,578 | 0,005 |

The developed algorithm is characterized by an approach similar to the routing algorithm proposed independently for recursive circulants in [19]. Due to the fact that the algorithm, obtained in this work, was developed on the basis of the features of the structure of multiplicative circulants, which are a articular case of recursive circulants, it is characterized by a lower computational complexity, but it can only be used for this class of circulants.

## 5. Approbation of algorithm operation

Testing of developed algorithms was performed on FPGA Cyclone V 5CGXFC9A6U19I7 form Intel FPGA (Altera). The description of the routers was done in Verilog [24]. Testing [25] was conducted in two stages. At the first stage, the data on the occupied chip resources was obtained for one router and the network as a whole for those considered earlier (Figure 3a, Figure 3b).

Based on the data obtained, it can be concluded that the major cause for increase in the router-occupied chip resources is the number of multiplicative network generatrices.. With an increase in number of generatrices in the structure of the circulant, the number of mathematical operations for their verification, as well as memory for storing the values of the generators, increases.

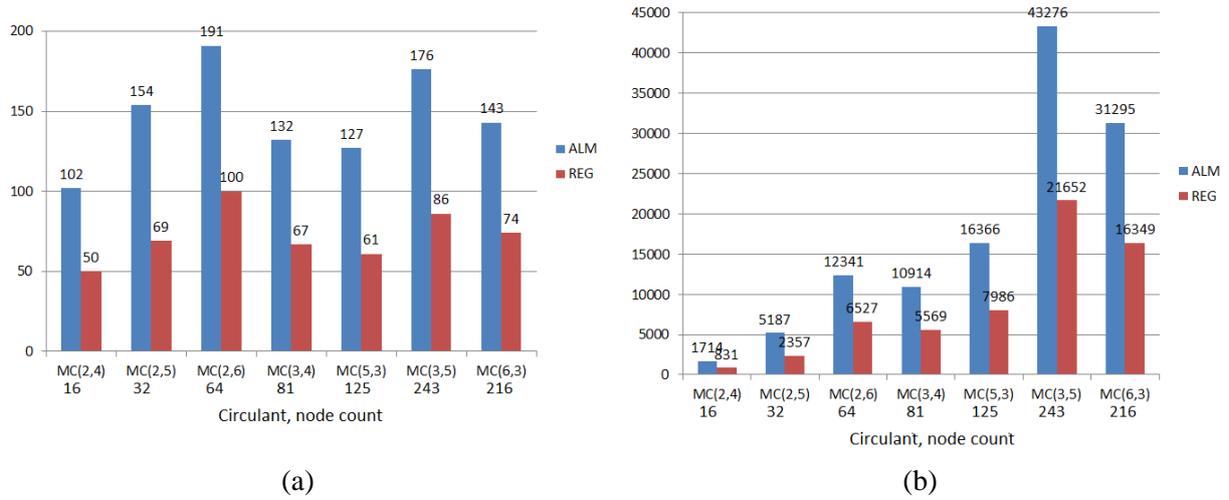

(a) (b)

**Figure 3.** Number of resources occupied by the router (a) and by the network (b) on a chip.

At the second stage, for comparison with other families of circulants, the circulants, which can be described in several representations, were chosen. The comparison was made between multiplicative circulants with the number of generatrices equal to 2, and ring circulants. The number of network nodes was formed as a natural number in degree 2. As a result of RTL synthesis, the data on the number of ALM blocks and registers (Table 3, Figure 4a, b) was obtained.

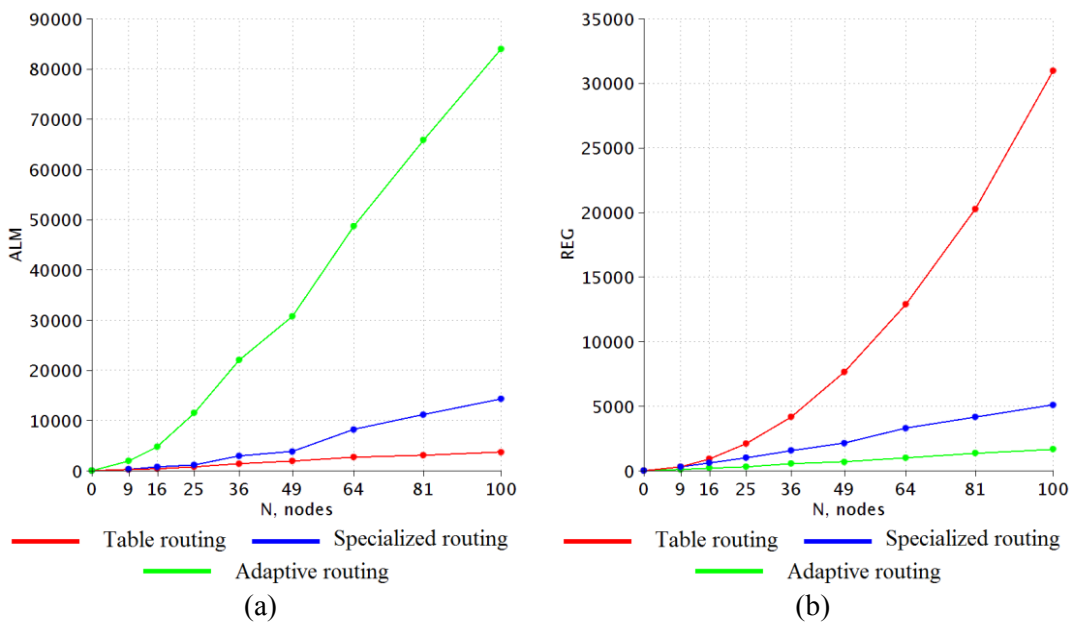

(a) (b)

**Figure 4.** Dependence of used ALM chip blocks (a) and REG (b) on the routing algorithm.

Table 3. Resources occupied chip blocks for the whole network.

| Circulant | Number of nodes | ALM (Table alg) | REG (Table alg) | ALM (Ring) | REG (Ring) | ALM (MC) | REG (MC) |
|---|---|---|---|---|---|---|---|
| $MC(3,2)$ | 9 | 205 | 292 | 1933 | 81 | 275 | 293 |
| $MC(4,2)$ | 16 | 390 | 912 | 4737 | 210 | 732 | 605 |
| $MC(5,2)$ | 25 | 753 | 2089 | 11491 | 301 | 1121 | 1000 |
| $MC(6,2)$ | 36 | 1350 | 4158 | 22025 | 546 | 2906 | 1558 |
| $MC(7,2)$ | 49 | 1909 | 7637 | 30709 | 664 | 3774 | 2118 |
| $MC(8,2)$ | 64 | 2686 | 12879 | 48671 | 988 | 8157 | 3275 |
| $MC(9,2)$ | 81 | 3055 | 20249 | 65825 | 1356 | 11121 | 4151 |
| $MC(10,2)$ | 100 | 3644 | 30956 | 83938 | 1662 | 14278 | 5076 |

The columns "ALM (Table alg.)" and "REG (Table alg." represent the results of tabular routing, the columns "ALM (Ring)" and "REG (Ring)" show the results of the algorithm for describing the graph as a circular circulant [26], and in the columns "ALM (MC)" and "REG (MC)" – for multiplicative circulant. The graphs show that the rate of increase in the use of ALM blocks in the algorithm for describing the network as a multiplicative circulant is much lower than that of the algorithm for describing the network as a ring circulant. At the same time, the rate of increase in the use of registers in the algorithm, proposed for the multiplicative circulant, is slightly higher than that of the algorithm for the ring circulant. Given the fact that the logical resources of the chip are much smaller than the registers, it can be concluded that the use of multiplicative circulants is more effective than the ring ones with an equal number of generatrices.

## 6. Conclusion

In the conditions of inconsistency of the characteristics of common topologies to the requirements of modern networks and in view of the need to search for alternative options for building networks, it is suggested to consider as a topology a special kind of graphs – multiplicative circulants. Strict rules for the formation of the circulant structure impose restrictions on the number of nodes in the network, and make the topology data easily scalable to larger networks and allow improving such important characteristics as diameter and average network distance in comparison with classical topologies.

For multiplicative circulants, standard shortest path search algorithms are applicable, and in this paper we propose the address part of a packet to reduce its size when static routing. At the same time, most of the standard shortest path search algorithms are built around the graph; that is why with increasing complexity of the circulant, the running time of the algorithm dramatically increases. Taking into account the peculiarities of the structure of multiplicative circulants, it was developed a specialized algorithm for this class of circulants, which makes it possible to considerably simplify the structure of address part of the packet and to reduce the time for finding the optimal path in the graph.

**Acknowledgment**
The publication was prepared within the framework of the Academic Fund Program at the National Research University Higher School of Economics (HSE) in 2018-2019 (grant № 18-01-0074) and by the Russian Academic Excellence Project "5-100".

**References**
[1] Nychis G, Fallin C and Moscibroda T 2010 Next Generation On-Chip Networks: What Kind of Congestion Control Do We Need? *Hotnets-IX: Proceedings of the 9th ACM SIGCOMM Workshop on Hot Topics in Networks* (Monterey, California: ACM).
[2] Paul J, Stechele W and Oechslein B 2015 Resource-awareness on heterogeneous MPSoCs for image processing *Journal of Systems Architecture V. 61, I. 10.*(Elsevier) pp 668-680.
[3] Abdelfattah M S, Bitar A and Betz V 2017 Design and applications for embedded networks-on-chip on FPGAs *IEEE Transactions on Computers V. 66, I. 6.* pp 1008-1021.

[4] Deb D, Jose J and Das S 2019 Cost effective routing techniques in 2D mesh NoC using on-chip transmission lines *Journal of Parallel and Distributed Computing V. 123* (Elsevier) pp 118-129.
[5] Ansari A Q, Ansari M R and Khan M A 2016 Modified quadrant-based routing algorithm for 3D Torus Network-on-Chip architecture *Perspectives in Science V. 8,* pp 718-721.
[6] Marvasti M B and Szymanski T H 2012 The performance of hypermesh NoCs in FPGAs *IEEE 30th International Conference on Computer Design (ICCD)* (Montreal, Canada) pp 492-493.
[7] Bishnoi R, Kumar P and Laxmi V 2014 Distributed adaptive routing for spidergon NoC *18th International Symposium on VLSI Design and Test* (IEEE) pp 1-6.
[8] Dally W J and Towles B 2004 Principles and practices of interconnection networks. (Elsevier). p 550.
[9] Monahova Je A 2011 Strukturnye i kommunikativnye svojstva cirkuljantnyh setej *Prikladnaja diskretnaja matematika No. 3(13)* pp 92–115.
[10] Intel Corporation 'Single-chip cloud computer', 2009 [Online]. Available: https://www.intel.ru/content/dam/www/public/us/en/documents/technology-briefs/intel-labs-single-chip-cloud-overview-paper.pdf (Accessed: 20-Oct-2018).
[11] Intel Corporation 'Intel's Teraflops Research Chip' [Online]. Available: http://download.intel.com/pressroom/kits/Teraflops/Teraflops_Research_Chip_Overview.pdf (Accessed: 20-Oct-2018).
[12] Elspas B and Turner J 1970 Graphs with circulant adjacency matrices *Journal of Combinatorial Theory V. 9. Iss. 3.* (California: Stanford Research Institute / Elsevier) pp 297–307.
[13] Boesch F and Tindell R 1984 Circulants and their connectivities *Journal of Graph Theory. V. 8. I. 4.* (New Jersey: Stevens Institute of Technology) pp 487–499.
[14] Monahova Je A 2010 Mul'tiplikativnye cirkuljantnye seti // *Diskretnyj analiz i issledovanie operacij V. 17. No. 5.* pp 56–66.
[15] Stojmenovic I 1997 Multiplicative circulant networks. Topological properties and communication algorithms *Discrete Applied Mathimatics N. 77* (Ottava: University of Ottawa / Elsevier) pp 281–305.
[16] Wong C K and Coppersmith D 1974 A combinatorical problem related to multimodule memory organizations *Journal of the ACM V. 21, I. 3* (New York / ACM) pp 392–402.
[17] Arno S and Wheeler F S 1993 Signed digit representations of minimal Hamming weight *IEEE Transactions on Computers. V. 42. I. 8.* pp 1007–1010.
[18] Suboh S, Bakhouya M and Gaber J 2008 An interconnection architecture for network-on-chip systems *Telecommunication Systems V. 37. I. 1–3* pp 137–144.
[19] Park J H and Chwa K Y 1994 Recursive Circulant: a New Topology for Multicomputer Networks *Proceedings of the International Symposium on Parallel Architectures, Algorithms and Networks (ISPAN)*, (Kanazawa, Japan: IEEE Computer Society Press) p 73–80.
[20] Bundy A and Wallen L 1984 Breadth-First Search // *Catalogue of Artificial Intelligence Tools*. (Heidelberg / Springer) pp 13.
[21] Crespelle C and Gambette P 2010 Unrestricted and complete Breadth-First Search of trapezoid graphs in O(n) time *Information Processing Letters V 110 I 12-13 pp 497–502.*
[22] Gabis A.B and Koudil M 2016 NoC routing protocols – objective-based classification *Journal of Systems Architecture. V. 66–67* pp 14–32.
[23] Kuo-Shun D, Ho C T and Jyh-Jong T 1998 Matrix transpose on meshes with wormhole and XY routing *Discrete Applied Mathematics V. 83, I.1-3,* pp 41-59.
[24] IEEE Std, 2001 'IEEE Standard Verilog® Hardware Description Language'[Online] Available: http://www-inst.eecs.berkeley.edu/~cs150/fa06/Labs/verilog-ieee (Accessed 10- Oct-2018).
[25] Lezhnev E V 'NoCs with circulant toplogy comparison' [Online] Available: https://github.com/EvgenijLezhnev/NoC-Comparison (Accessed 13-Oct-2018).
[26] Romanov A Yu Development of routing algorithms in networks-on-chip based on ring circulant topologies. In Press.